\documentclass[prb,
                 twocolumn,
                 preprintnumbers,
                amsmath,
                amssymb,
                showpacs,
                superscriptaddress]{revtex4}
\usepackage{graphicx} 
\usepackage{dcolumn}  
\usepackage{bm}       
\usepackage{amsfonts}
\usepackage{dsfont}
\usepackage{csquotes}
\usepackage{comment}

\usepackage{ulem}
\usepackage{color}

\begin{document}

\title{High-Harmonic Generation from Monolayer and Bilayer Graphene}

\author{M. S. Mrudul}
\affiliation{%
Department of Physics, Indian Institute of Technology Bombay,
           Powai, Mumbai 400076, India }

\author{Gopal Dixit}
\email[]{gdixit@phy.iitb.ac.in}
\affiliation{%
Department of Physics, Indian Institute of Technology Bombay, Powai, Mumbai 400076, India }

\date{\today}


\begin{abstract}
High-harmonic generation (HHG) in solids is an emerging method to probe ultrafast electron dynamics in solids at
attosecond timescale. In this work, we study HHG from  
monolayer and bilayer graphene. 
Bilayer graphenes with  AA and AB stacking are considered in this work.
It is found that the monolayer and bilayer graphenes exhibit significantly different harmonic spectra. 
The difference in the spectra is attributed to the interlayer coupling between the two layers. 
Also, the intraband and interband contributions to the total harmonic spectrum play 
a significant role. 
Moreover, interesting polarization and ellipticity dependence are noticed in total harmonic spectrum for  monolayer and bilayer graphene. 

\end{abstract}

\maketitle 
\section{Introduction}
Thanks to the technological advances that have enabled us to study strong-field driven processes in solids. 
Strong-field driven electron dynamics opens an avenue to control and understanding 
carrier dynamics on  attosecond timescale~\cite{kruchinin2018colloquium}. 
High-harmonic generation (HHG) is one such strong-field driven process 
in which radiation with integer multiples of the incident driving frequency is emitted. 
After the pioneering work of Ghimire \textit{et al.}~\cite{ghimire2011observation}, 
HHG in solids has opened the door to study 
the electronic structure and its dynamics with the characteristic timescale in solids~\cite{kruchinin2018colloquium, ghimire2019,catoire2018wannier}.  
In recent years, HHG in solids has been employed to explore several exciting processes such as 
band structure 
tomography~\cite{vampa2015all, lanin2017mapping, tancogne2017impact}, 
probing  the dynamics of the defects in solids~\cite{mrudul2020high, pattanayak2020influence}, the  
realization of petahertz current in solids~\cite{luu2015extreme, garg2016multi}, and
Bloch oscillations~\cite{schubert2014sub, mcdonald2015interband}. 
Moreover,  HHG in solids offers an attractive all-solid-state compact optical device to obtain coherent and bright attosecond pulses in the extreme ultraviolet energy regime~\cite{kruchinin2018colloquium, ghimire2019, nourbakhsh2020high}.

The realization of an atomically-thin monolayer graphene has catalyzed  a series of breakthroughs in 
fundamental and applied sciences~\cite{geim2009graphene}. 
Graphene shows unusual electronic and optical properties in comparison  to its bulk counterpart~\cite{novoselov2004electric}. 
The unique electronic structure of graphene exhibits  varieties of nonlinear optical processes~\cite{avetissian2016coherent, hendry2010coherent, kumar2013third}. 
HHG from monolayer and  few-layer graphenes has been extensively studied in the past~\cite{hafez2018extremely, avetissian2018impact, chizhova2017high, al2014high, yoshikawa2017high, zurron2019optical, taucer2017nonperturbative, liu2018driving, chen2019circularly, mikhailov2007non, gupta2003generation, jimenez2020light,sato2021high}. 
The underlying mechanism of HHG in graphene~\cite{zurron2018theory} was 
different from the one explained using a two-band model by Vampa \textit{et al.}~\cite{vampa2014theoretical}. 
The intraband current from the linear band-dispersion of graphene was expected to be the dominating mechanism~\cite{mikhailov2007non, gupta2003generation}. 
This is a consequence of the highly non-parabolic nature of the energy-bands~\cite{ghimire2011observation}. 
In contrast to this prediction, the interband and intraband mechanisms in graphene are found to be coupled ~\cite{taucer2017nonperturbative, liu2018driving, al2014high,sato2021high} except for low intensity driving fields~\cite{al2014high}. Vanishing band-gap and diverging dipole matrix elements near 
Dirac points lead to strong interband mixing of valence and conduction bands in graphene~\cite{zurron2018theory,kelardeh2015graphene}. The ellipticity dependence of HHG from graphene has been  observed experimentally~\cite{yoshikawa2017high, taucer2017nonperturbative} as well as discussed theoretically~\cite{yoshikawa2017high, taucer2017nonperturbative, liu2018driving, zurron2019optical}. 
Taucer \textit{et al.}~\cite{taucer2017nonperturbative} have demonstrated  that 
the ellipticity dependence of the harmonics in graphene is atom like,  
whereas a higher harmonic yield for a particular ellipticity was observed by Yoshikawa \textit{et al.}~\cite{yoshikawa2017high}. 
The anomalous ellipticity dependence was attributed to the strong-field interaction in the semi-metal regime~\cite{tamaya2016diabatic, yoshikawa2017high}.

Along with monolayer graphene, bilayer graphene is also attractive due to its interesting optical response~\cite{yan2012tunable}. A bilayer graphene can be made by stacking another layer of the monolayer graphene on top of the first.   
Three suitable configurations  of the bilayer graphene are possible: 
(a) AA stacking in which the second layer is placed exactly on top of the first layer; (b) 
AB stacking in which the B atom of the upper layer is placed on the top of the A atom of the lower layer, whereas the other type of atom occupies the center of the hexagon; 
and (c) twisted bilayer in which the upper layer is rotated by an angle with respect to the lower layer. 
AB stacking, also known as the Bernel stacking, is the one that is a more energetically stable structure and mostly realized in experiments [see Fig.~\ref{fig1}(b)]~\cite{rozhkov2016electronic, mccann2013electronic}. Recently, electron dynamics~\cite{ulstrup2014ultrafast,kumar2020bilayer, hipolito2018nonlinear} and valley polarization in bilayer graphene on ultrafast timescale have been
 discussed theoretically~\cite{kumar2020ultrafast}. 
Avetissian \textit{et al.} have discussed the role of the multiphoton resonant excitations in HHG for AB stacked bilayer graphene~\cite{avetissian2013multiphoton}.  
Moreover, HHG from a twisted bilayer graphene has been explored recently~\cite{PhysRevResearch.2.032015}. 
However, the comparison of HHG from monolayer and bilayer graphenes; and a thorough investigation of the role of interlayer coupling are unexplored.

In this work, we investigate HHG from monolayer and bilayer graphenes with 
AA and AB configurations. 
Moreover, 
the roles of interband and intraband contributions are investigated in both  cases. 
The role of interlayer coupling in HHG from bilayer graphene is investigated. 
Furthermore, polarization and ellipticity dependences of the HHG are discussed. 
This paper is organized as follows: The theoretical model and numerical methods are presented in Sec. II. 
Sec. III presents the results and discussion of our numerical simulations, and the conclusions are presented in Sec. IV.

\section{Numerical Methods}\label{section:2}

\begin{figure}[h!]
	\includegraphics[width=\linewidth]{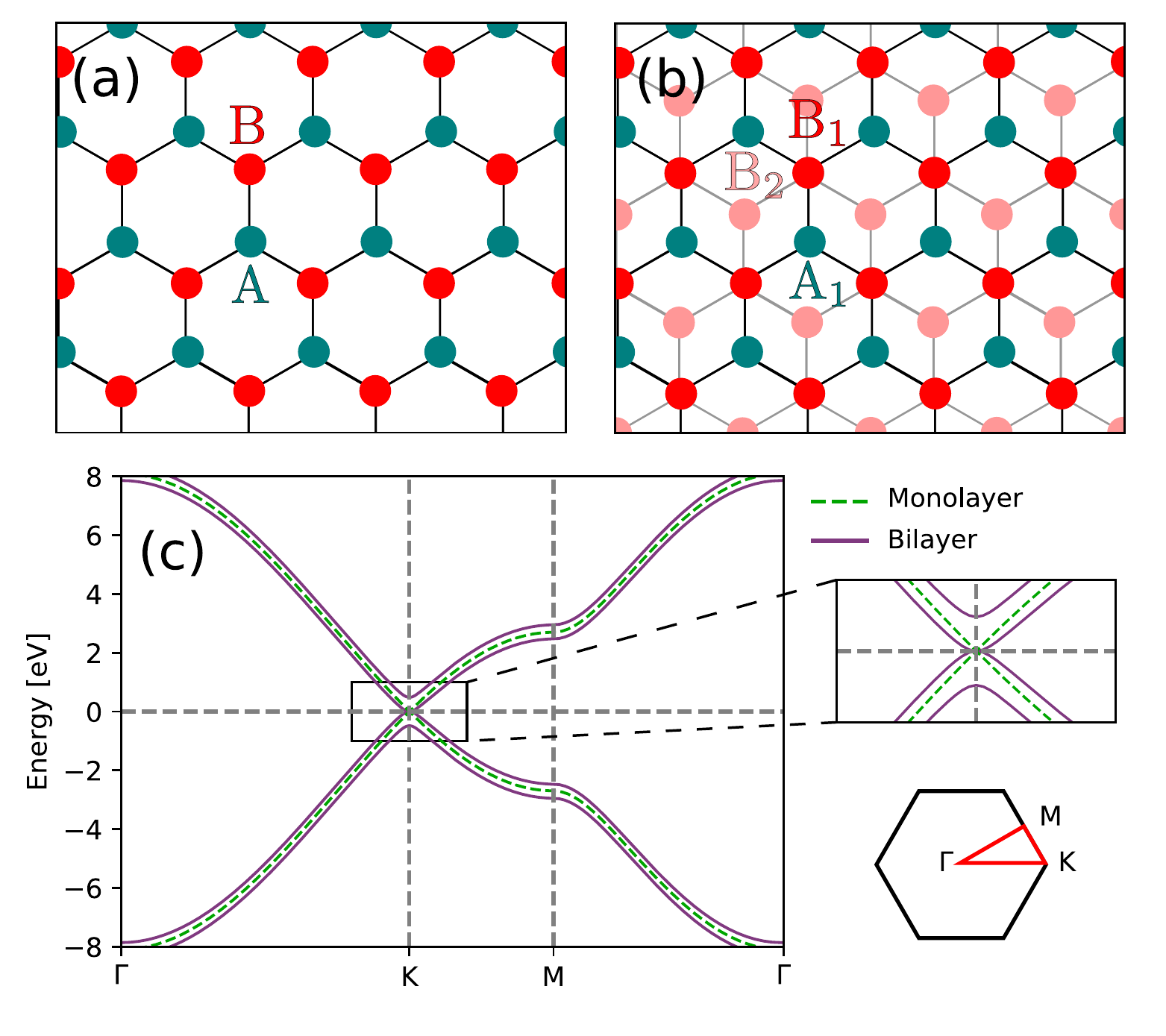}
	\caption{Monolayer and bilayer graphene (AB-stacked). (a) and (b) 
	The top-view of the monolayer and bilayer graphenes, respectively. 
	The carbon atoms are arranged in a honeycomb lattice with two inequivalent carbon atoms (A and B). 
	In bilayer graphene, the B atoms of the top layer are placed on top of A atoms of the bottom layer. 
	(c)  The band-structure of the monolayer (green) and bilayer graphene (violet).} \label{fig1}
\end{figure}

The real-space lattice of monolayer graphene is shown in Fig.~\ref{fig1}a. 
Carbon atoms are arranged in a honeycomb lattice with a two-atom basis unit-cell. 
The A and B atoms in Fig.~\ref{fig1}(a) represent two inequivalent carbon atoms in the unit-cell. The lattice parameter of graphene is equal to  2.46 \AA. Nearest-neighbour tight-binding approximation is implemented by only considering the p$_z$ orbitals of the carbon atoms. The Hamiltonian for monolayer graphene is defined as
\begin{equation}
 	\hat{H}_{0} = -t_0 f(\textbf{k})\hat{a}^{\dagger}_k \hat{b}_k + \textrm{H.c.}
\end{equation}
Here, $\hat{a}_k$, $\hat{b}_k$ are, respectively, the annihilation operators associated with A and B types of the atoms in the unit-cell. 
$f(\textbf{k})$ is defined as $f(\textbf{k}) = \sum_i e^{i\textbf{k}\cdot \delta_i}$, where $\delta_i$ is  the nearest neighbour vector.  
A nearest-neighbour in-plane hopping energy $t_0$  of 2.7 eV is used~\cite{reich2002tight,trambly2010localization,moon2012energy}.  
The eigenvalues of the Hamiltonian are given by $\mathcal{E}(\textbf{k}) = \pm t_{0}|f(\textbf{k})|$.

Similarly, the Hamiltonian for AB-stacked bilayer graphene can be defined as
\begin{equation}
	\hat{H}_{AB} = -t_0 f(\textbf{k}) \left[\hat{a}^\dagger_{1k}\hat{b}_{1k} + \hat{a}^\dagger_{2k}\hat{b}_{2k}   \right] + t_\perp \hat{a}^\dagger_{2k}\hat{b}_{1k} + \textrm{H.c.}
\end{equation}
Here, 1 and 2 denote the carbon atoms in the upper and lower layers, respectively. 
An inter-plane hopping energy $t_\perp$ of 0.48 eV is used for an inter-layer separation equal to  3.35 \AA~\cite{trambly2010localization,moon2012energy}. 
The band-structure for the bilayer graphene is given as $\mathcal{E}(\textbf{k}) =  [\pm t_\perp \pm \sqrt{4 |f(\textbf{k})|^2 t_0^2 + t_\perp^2}]/2 .$

Figure~\ref{fig1}(c) presents the energy band-structure of both monolayer and bilayer graphenes. 
The band-structure of monolayer graphene has zero band-gap and linear dispersion near 
two points, known as $\bf{K}$-points, in the Brillouin zone (BZ). 
On the other hand, bilayer graphene near $\bf{K}$-points shows a quadratic dispersion. 
Due to the zero band-gap nature, both monolayer and bilayer graphene are semi-metals. 
Here, electron-hole symmetry is considered by neglecting higher-order hopping and overlap of the orbitals.

Semiconductor Bloch equation in Houston basis is solved as~\cite{houston1940acceleration, krieger1986time, floss2018ab}  

\begin{widetext}
\begin{equation}
  i\frac{d}{dt} \rho _{mn}^{\textbf{k}}  =   \mathcal{E}_{mn}^{\textbf{k} + \textbf{A}(t)}\rho_{mn}^{\textbf{k}} + i\frac{\tilde{\delta}_{mn}}{T_2} \rho_{mn}^{\textbf{k}}
	- 
		  \textbf{F}(t)\cdot\sum_l \left(\textbf{d}_{ml}^{~\textbf{k}+\textbf{A}(t)}\rho_{ln}^{\textbf{k}} - \textbf{d}_{ln}^{~\textbf{k}+\textbf{A}(t)}\rho_{ml}^{\textbf{k}}  \right) ,
\end{equation}
\end{widetext}
where $\textbf{F}(t)$ and $\textbf{A}(t)$ are  the electric field and vector potential of the driving laser field, respectively,
and are related as $\textbf{F}(t) = - d\textbf{A}(t)/ dt$. 
$\mathcal{E}^{\textbf{k}}_{mn}$ and  $\textbf{d}^{~\textbf{k}}_{mn}$ are the band-gap energy and the dipole-matrix elements between $m$ and $n$ bands, respectively.  
Here, $\tilde{\delta}_{mn}$ is defined as (1-$\delta_{mn}$);
$\textbf{d}_{mn}$ are calculated as $\textbf{d}_{mn} = -i\left\langle u_{m,\textbf{k}}\right|\nabla_{\textbf{k}}\left|u_{n,\textbf{k}} \right\rangle$, where $\left|u_{m,\textbf{k}}\right\rangle$ is the periodic part of the wavefunction.
A phenomenological term accounting for the decoherence is added, with a constant dephasing time $T_2$. 
For HHG from monolayer graphene, a dephasing time within the range of 2 fs to 35 fs has been used in the past~\cite{sato2021high,liu2018driving,taucer2017nonperturbative}. 
Moreover, a detailed investigation about dephasing time dependence on HHG from monolayer graphene has been discussed in Ref.~\cite{liu2018driving}. 
In this work,  dephasing time of 10 fs is  considered for monolayer and bilayer graphene.

 The total current at any $\bf{k}$-point in the BZ is calculated as
\begin{equation}
\begin{split}
\textbf{J}(\textbf{k}, t) &= \sum_{m,n} \rho_{mn}^{\textbf{k}}(t)~\textbf{p}_{nm}^{~\textbf{k}+\textbf{A}(t)} \\
&= \sum_{m\neq n} \rho_{mn}^{\textbf{k}}(t)~\textbf{p}_{nm}^{~\textbf{k}+\textbf{A}(t)} + \sum_{m=n} \rho_{mn}^{\textbf{k}}(t)~\textbf{p}_{nm}^{~\textbf{k}+\textbf{A}(t)}\\
&= \textbf{J}_{inter}(\textbf{k}, t) + \textbf{J}_{intra}(\textbf{k}, t). 
\end{split}
\end{equation}
Here, $\textbf{p}^{\textbf{k}}_{nm}$ are the momentum matrix elements, and $\textbf{J}_{inter}(\textbf{k}, t)$ and $\textbf{J}_{intra}(\textbf{k}, t)$ are interband and intraband contributions to the total current, respectively. The high-order harmonic spectrum is 
determined from the Fourier-transform of the time-derivative of the current as
\begin{equation}
\mathcal{I}(\omega) = \left|\mathcal{FT}\left(\frac{d}{dt} \left[\int_{BZ} \textbf{J}(\textbf{k},t)~\rm d\textbf{k} \right]\right) \right|^2 .
\end{equation}

In the present work, driving laser pulse with an intensity of 1$\times$10$^{11}$ W/cm$^2$ and 
wavelength of 3.2 $\mu$m is used. The laser pulse is  eight-cycles in duration with a sin-squared  envelope.  
The intensity of the driving pulse is much below the damage threshold and lower than the one used in experimental HHG from graphene~\cite{yoshikawa2017high, taucer2017nonperturbative}. The same parameters  of the driving laser pulse are used throughout unless stated otherwise.

\section{Results and Discussions}

\begin{figure}[h!]
	\includegraphics[width=\linewidth]{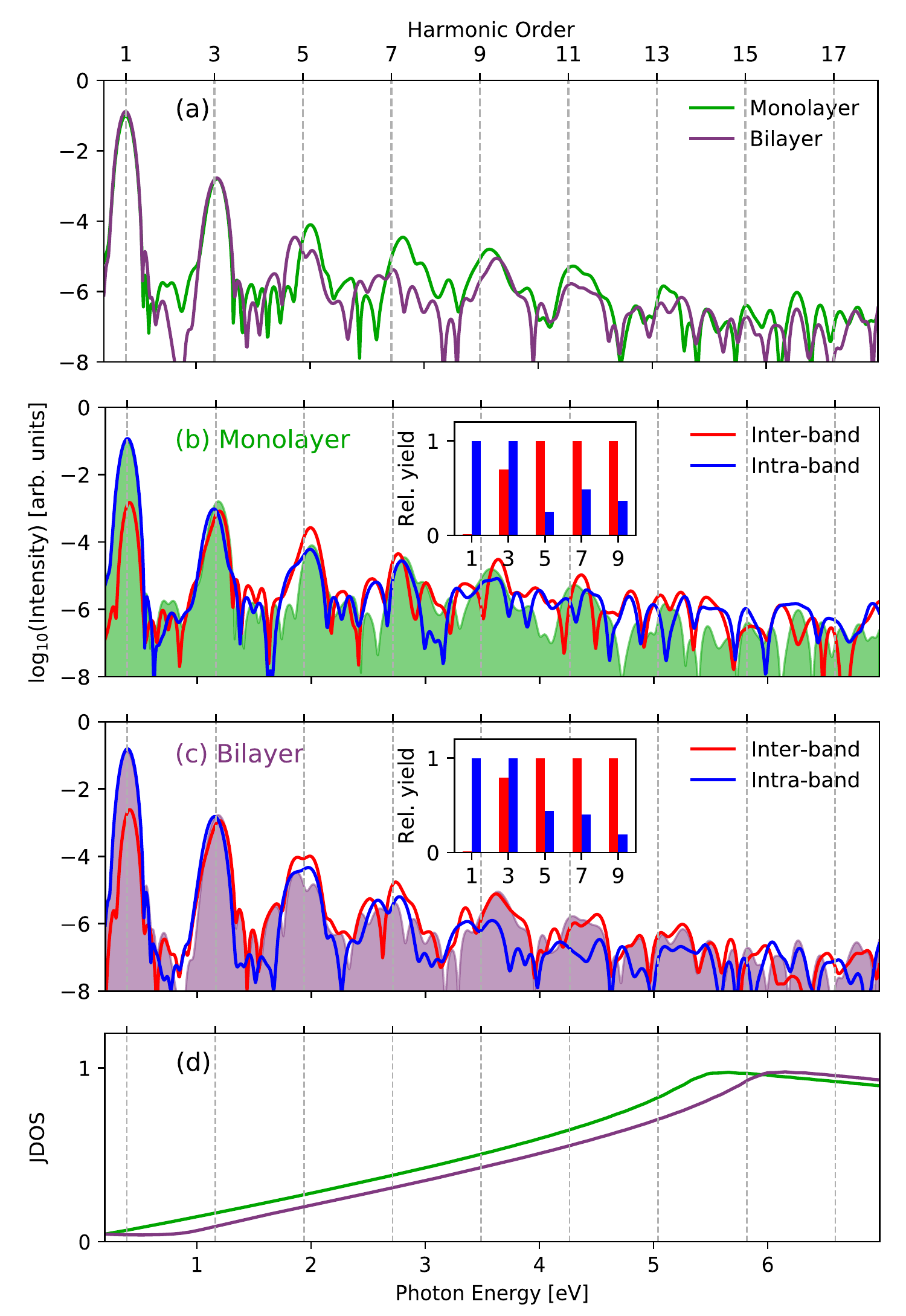}
	\caption[High harmonic spectrum]
	{(a) High harmonic spectrum of the monolayer (green) and bilayer (violet)   
	graphene with AB stacking. A linearly polarized pulse along the $\Gamma$-K direction is considered here. The high-harmonic intensity is normalized to the total number of electrons in 
	monolayer and bilayer graphenes. The interband and intraband contributions to the high-harmonic spectrum for 
	(b) monolayer and (c) bilayer graphenes. The total harmonic spectrum is also plotted for the reference. 
	The relative harmonic yield (integrated) for different orders from interband and intraband contributions is plotted in the insets of (b) and (c), respectively. (d) The normalized optical joint density of states (JDOS) of monolayer (green) and bilayer (violet) graphenes.} \label{fig2}
\end{figure}

Figure~\ref{fig2} presents the HHG spectra of monolayer graphene and its comparison with the spectra of the  bilayer graphene for a  linearly polarized laser pulse having 
polarized along the $x$-axis ($\Gamma$-K in the BZ).
Here, AB stacking of bilayer graphene is considered. The intensity of  the HHG spectra is normalized 
with respect to  the total number of electrons in monolayer and bilayer graphenes. 
It is  apparent that the third harmonic (H3) is matching well  in both cases. 
However, harmonics higher than H3 show significantly different behaviour
as the interlayer coupling between the two layers plays a meaningful role.

The contributions of  interband and intraband to the total harmonic spectra for monolayer and bilayer graphene are shown in Figs.~\ref{fig2}(b) and 2(c), respectively. 
Both  contributions play a  strong role to the total spectra as reflected from the figure. 
A strong interplay of interband and intraband contributions was reported for monolayer graphene~\cite{taucer2017nonperturbative, liu2018driving, al2014high}. 
Unlike the wide band-gap semiconductors~\cite{wu2015high},  the interband and intraband transitions take place at the same energy scales for both monolayer and bilayer graphenes~\cite{stroucken2011optical} due to the vanishing band-gap. 
The relative  (integrated) harmonic yield from interband and intraband contributions is plotted in the insets of Figs.~\ref{fig2}(b) and 2(c). Here, intraband contribution dominates upto H3, 
whereas interband contribution dominates for fifth (H5) and higher order harmonics for both monolayer and bilayer graphenes. The enhanced contributions from interband transitions at higher orders can be attributed to the increased joint density of states at higher energies as shown in Fig.~\ref{fig2}(d).

Also, as the low energy band-structures are different for monolayer and bilayer graphenes  [Fig.~\ref{fig1}(c)], 
the nature of harmonic spectra is not obvious from the band-structure point of view. 
To have a better understanding of the underlying mechanism of the 
harmonic generation in both  cases,  
the role of the interlayer coupling in HHG is discussed in the next subsection.

\subsection{Role of interlayer coupling in  HHG}

\begin{figure}
	\centering
	\includegraphics[width=\linewidth]{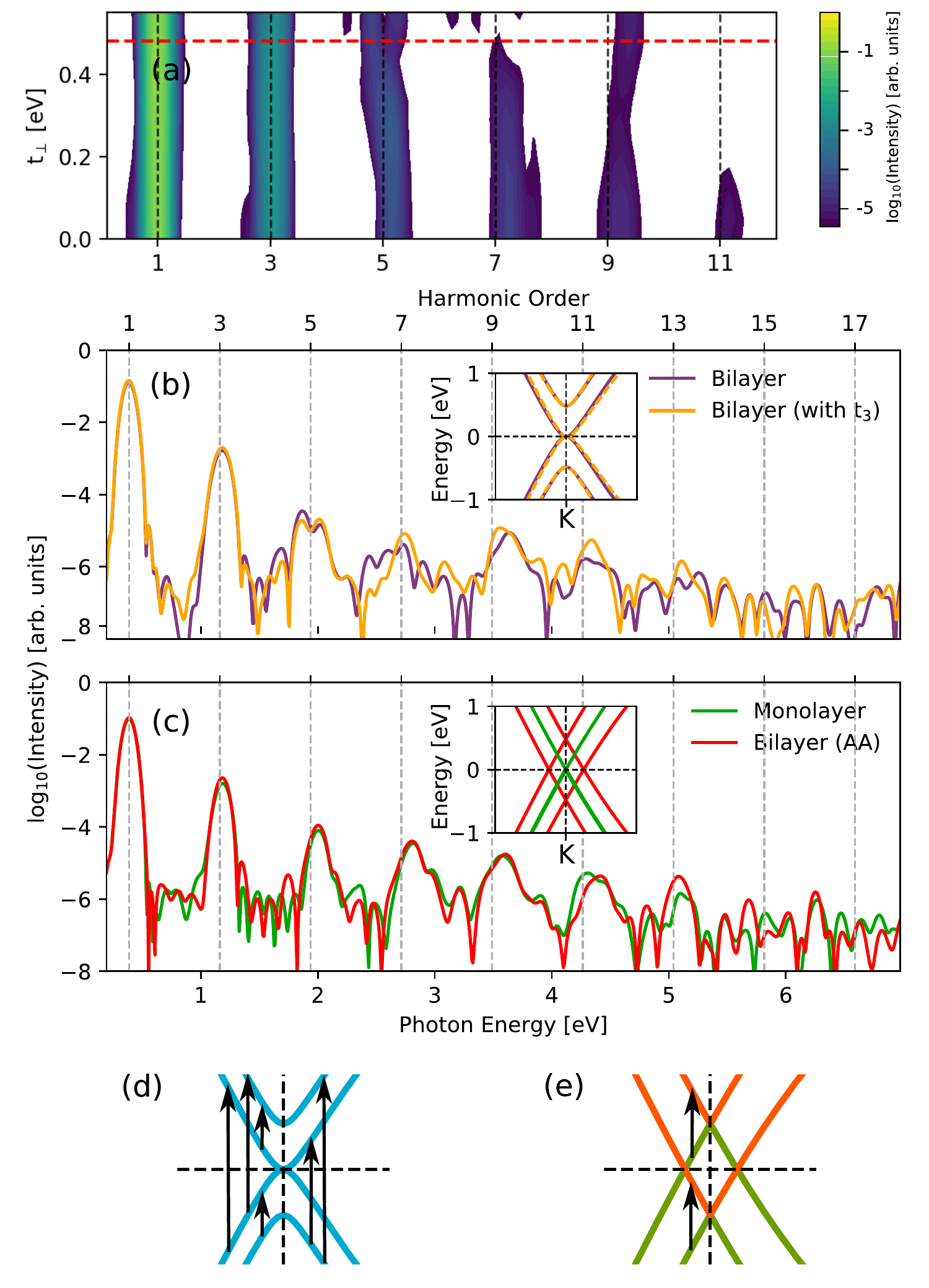}
	\caption[]{(a) High harmonic spectrum  of bilayer graphene with AB stacking as a function of interlayer coupling ($t_{\perp}$), where $t_{\perp}$ = 0 corresponds to the HHG from monolayer graphene. The red dashed line corresponds to the actual value of $t_{\perp}$ used in the calculations. (b) HHG from bilayer graphene (in AB stacking) with $t_{\perp}$ (B$_1$-A$_2$) coupling only (violet)  and with both $t_{\perp}$ and $t_3$ (A$_1$-B$_2$) coupling (orange). (c) HHG from bilayer graphene with AA stacking (red) and monolayer graphene (green). 
	In (b) and (c), the band-structures near the K-point are shown in the inset. (d) and (e) show the non-zero momentum matrix elements in AB and AA stacked bilayer graphene, respectively.}
	\label{fig:fig5}
\end{figure}

To understand how the interlayer coupling between two layers affects the harmonic spectrum of bilayer graphene, 
the harmonic spectrum as a function of interlayer coupling strength ($t_{\perp}$) is shown in Fig.~\ref{fig:fig5}(a). 
Reducing the interlayer coupling strength is equivalent to moving the two layers of graphene
farther apart. 
The red dashed line in Fig.~\ref{fig:fig5}(a) corresponds to the interlayer coupling used in simulations shown in Fig.~\ref{fig2}. 
It is evident from Fig.~\ref{fig:fig5}(a) that  H5 and higher-order harmonics are changing with respect to $t_{\perp}$. 
Moreover,  different harmonic orders affected differently. 
Therefore,  harmonic orders are non-linear functions of interlayer coupling.

To explore further how different hopping terms affect the HHG in bilayer graphene, 
an additional hopping term,  $t_3$, between B atoms of the top layer and A atoms of the bottom layer is introduced. The modified Hamiltonian for AB-stacked bilayer graphene can be written as, 
\begin{equation}
\begin{split}
\hat{H}_{AB} =& -t_0 f(\textbf{k}) \left[\hat{a}^\dagger_{1k}\hat{b}_{1k} + \hat{a}^\dagger_{2k}\hat{b}_{2k}   \right] \\
&+ t_\perp \hat{a}^\dagger_{2k}\hat{b}_{1k}  -t_3 f^*(\textbf{k}) \hat{a}^\dagger_{1k}\hat{b}_{2k} + 
\textrm{H.c.}
\end{split}
\end{equation}
Here, a hopping energy $t_3$ of 0.3 eV is used~\cite{charlier1991first,min2007ab}.  
The corresponding harmonic spectrum is presented in Fig.~\ref{fig:fig5}(b). 
It is evident from the figure that the additional interlayer coupling $t_3$ affects all the harmonics higher than H3. 
It is apparent that the interlayer coupling has a strong role in determining the non-linear response of bilayer graphene.

Now let us discuss how HHG depends on different stacking configurations of the bilayer graphene. 
As discussed in the introduction, bilayer graphene can be realized in AA and AB stacking. 
AA stacking of  bilayer graphene is realized by stacking the monolayer precisely on  
top of the first layer. 
The top view of the AA-stacked bilayer looks exactly as a monolayer graphene [Fig.~\ref{fig1}(a)], 
where A$_1$ couples with A$_2$ and B$_1$ couples with B$_2$ with a coupling strength of $t_{\perp}$. 
The harmonic profile of the bilayer graphene with AA configuration matches well with the spectrum of monolayer graphene as presented in Fig.~\ref{fig:fig5}(c).

The band-structures near the $\bf{K}$-point for AB and AA stacked bilayer graphene are shown in the insets of Figs.~\ref{fig:fig5}(b) and 3(c), respectively.  
For AB-stacked bilayer graphene, a slight change in band-structure results in a significant change in the spectrum [see Fig.~\ref{fig:fig5}(b)]. 
On the other hand, for AA-stacked bilayer graphene, the difference in the band-structure is not reflected in the spectrum [see Fig.~\ref{fig:fig5}(c)].

A better understanding about the HHG mechanism can be deduced by considering the roles of the band-structure as well as the interband momentum-matrix elements.   
The energy-bands of the AA-stacked bilayer graphene within nearest neighbour tight-binding approximation are given by $\mathcal{E}(\textbf{k}) = \pm t_{\perp} \pm t_0 |f(\textbf{k})|$. 
This is equivalent to the shifted energy-bands of monolayer graphene by $\pm t_{\perp}$. 
Also the corresponding momentum matrix elements give non-zero values only for the pairs $t_{\perp} \pm t_0 |f(\textbf{k})|$ and $-t_{\perp} \pm t_0 |f(\textbf{k})|$ as shown in Fig.~\ref{fig:fig5}(e).  
On the other hand, in AB-stacked bilayer graphene, all pairs of bands have non zero momentum matrix elements near the $\bf{K}$-point as shown in Fig.~\ref{fig:fig5}(d). 
The similar band-dispersion and joint density-of-states compared to monolayer graphene results in similar harmonic spectrum in AA-stacked bilayer graphene. 
On the other hand, in bilayer graphene, an electron in the conduction band can recombine to a hole in 
any of the valence bands near the $\bf{K}$-points as shown in Fig~\ref{fig:fig5}(d). 
These different interband channels interfere and therefore generate 
the resulting harmonic spectrum for the AB-stacked bilayer graphene.

From here onward only bilayer graphene with AB stacking is considered, 
as the HHG spectra of the monolayer and bilayer graphenes with AA stacking  are the same.  
In the next subsections, we explore polarisation and ellipticity dependences of the HHG 
from monolayer and bilayer graphenes. 

\subsection{Polarization Dependence of the High-Harmonic Spectrum}

The vector potential corresponding to a linearly polarized laser pulse can be defined as
\begin{equation}
\textbf{A}(t) = A_0f(t) \cos(\omega t)\left[\cos(\theta)\hat{\textbf{e}}_x + \sin(\theta) \hat{\textbf{e}}_y\right].
\end{equation}
Here, $f(t)$ is the envelope function and $\theta$ is the angle between laser polarization and the 
$x$-axis ($\Gamma$-K in the BZ). 

\begin{figure}[h!]
	\includegraphics[width=\linewidth]{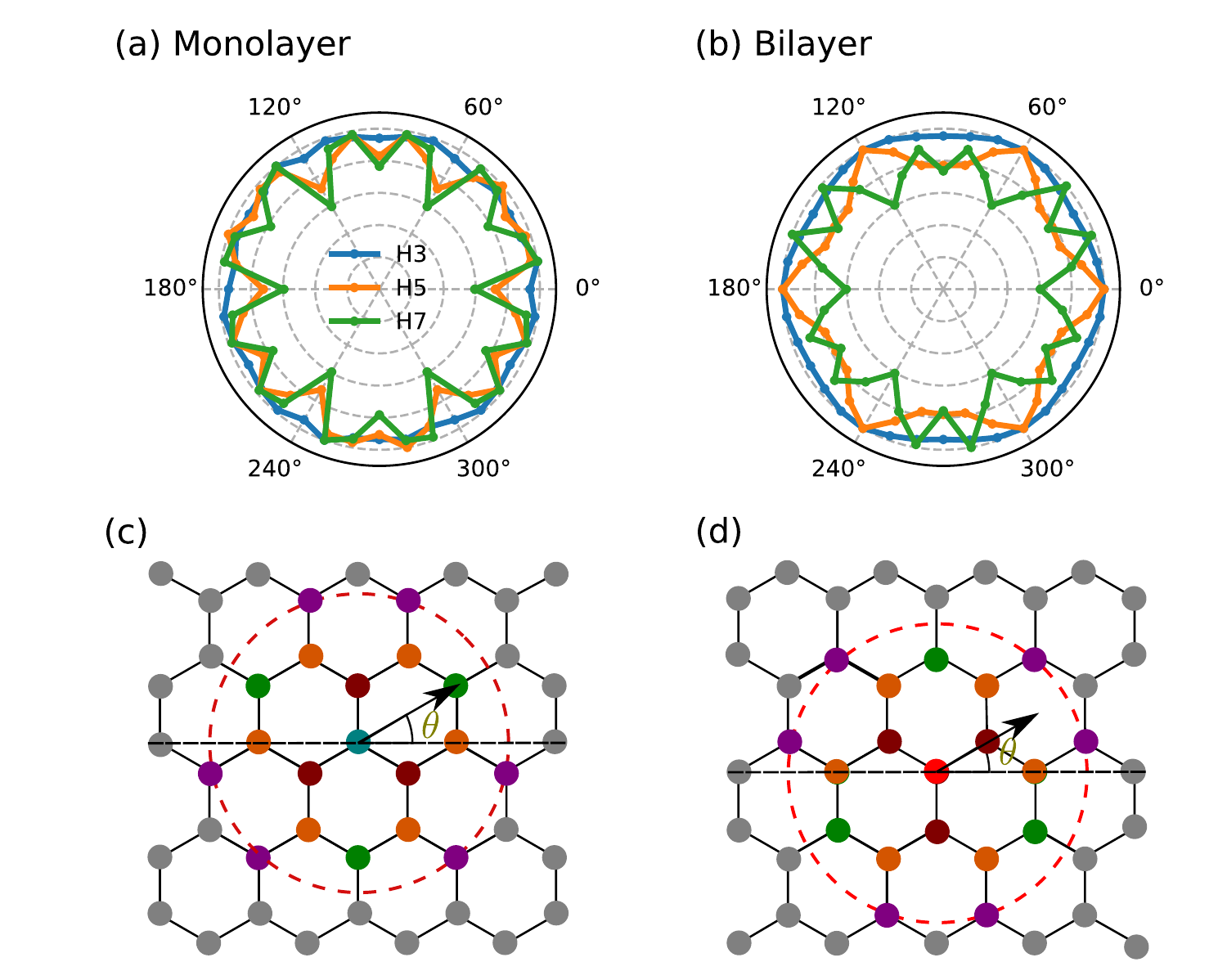}
	\caption{Polarization dependence of the normalized harmonic yield for (a) monolayer and (b) bilayer graphenes.
	 Here,  $\theta$ is  an angle between laser polarization and the $x$-axis along  $\Gamma$-K in the BZ.  
	 An illustration of the semi-classical real-space model with nearest neighbours of (e) A type and (d) B type carbon atoms. The first, second, third and fourth nearest  neighbours are shown using brown, orange, green and violet colours, respectively. } \label{fig3}.
\end{figure} 

The polarization-direction dependence of the harmonic yield for  monolayer and bilayer graphenes is presented 
in Figs.~\ref{fig3}(a) and \ref{fig3}(b), respectively. All the harmonics mimic the six-fold symmetry of the graphene lattice.  
As reflected from the figure, H3 exhibits no  significant polarization sensitivity 
for both monolayer and bilayer graphenes.  
The reason for this isotropic nature can be attributed to the  isotropic nature of the energy 
bands near $\bf{K}$-points. 
However, harmonics, higher than H3, show anisotropic behaviour in both   cases.  
Moreover, H5 of monolayer and bilayer graphenes shows different polarization dependence.
The harmonic yield is maximum for angles close to 15$^\circ$ and 45$^\circ$ in monolayer graphene.  

To understand the polarization dependence of the harmonic yield in monolayer graphene, we employ 
 a semiclassical explanation as proposed in Refs.~\cite{you2017anisotropic, pattanayak2019direct} by assuming that 
the interband transitions can be translated to a semi-classical real-space model~\cite{kruchinin2018colloquium,parks2020wannier}. 
One-to-one correspondence between interband transition 
and inter-atom hopping in graphene was shown by Stroucken \textit{et al.}~\cite{stroucken2011optical}. 
Here, we assume that an electron can hop between two atoms when the laser is polarized along a direction in which it connects the atoms. The contributions to the harmonic yield from different pairs of atoms drop significantly as the distance between the atoms increases.  
This is in principle governed by the inter-atom momentum matrix elements~\cite{stroucken2011optical}. 
By assuming a finite radius for atoms, farther atoms show sharper intensity peaks as a function of angle of polarization. 

Figures~\ref{fig3}(c) and 4(d)  show the nearest neighbours of A and B types of atoms in the unit-cell, respectively. 
Brown, orange, green and violet coloured atoms are, respectively, first, second, third and fourth neighbours. By only considering the nearest-neighbour hopping in graphene, we can see 
that the maximum yield should be for 30$^\circ$ ($\Gamma$-M direction). However, the maximum yield is near 15$^\circ$ and 45$^\circ$  as seen from  Fig.~\ref{fig3}(a). 
This means that the contributions up to the fourth nearest neighbours should be considered to explain the polarization dependence of H5 and seventh harmonic (H7) 
of monolayer graphene. 

In bilayer graphene, H7 follows the same qualitative behaviour as that of H5 and H7 of monolayer graphene [Fig.~\ref{fig3}(b)]. 
In contrast, H5 shows different behaviour and obeys the symmetry of the second nearest neighbour. It is clear from Fig.~\ref{fig:fig5}(d) that there are multiple paths for interband transitions for bilayer graphene. 
In bilayer graphene, interband transitions from different pairs of valence and conduction bands can contribute to a particular harmonic, 
and these different transitions interfere. 
This makes the mechanism of harmonic generation from monolayer and AB-stacked bilayer graphene different. 

It is important  to point out that the polarization dependence is sensitive  to the 
wavelength of the driving laser pulse.  
For longer wavelengths, electron dynamics occurs in the isotropic parts of the reciprocal space 
(close to $\mathbf{K}$-points), and as a result the harmonic spectrum can be entirely isotropic. 
We have confirmed  that the different symmetry observed for monolayer and bilayer graphenes is consistent with varying wavelengths of the driving laser (not shown here), 
and our explanation remains consistent.

\subsection{Ellipticity Dependance of the High-Harmonic Spectrum}

\begin{figure}[h!]
	\includegraphics[width=\linewidth]{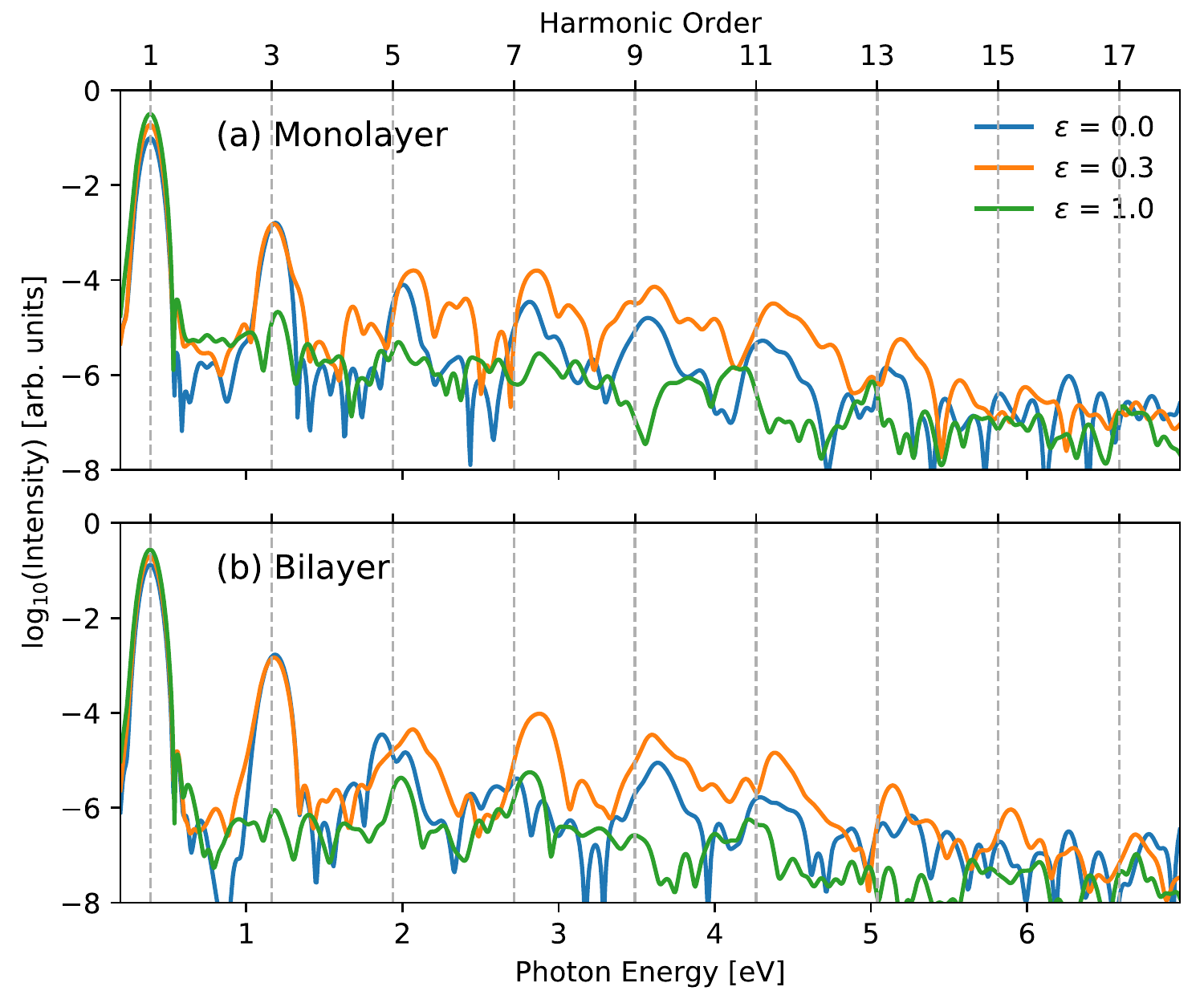}
	\caption{High harmonic  spectrum for (a) monolayer and (b) bilayer graphenes  
	for different ellipticities of the driving laser pulse. 
	Here, $\epsilon$ = 0 corresponds to a linearly polarized pulse 
	and $\epsilon$ = 1 corresponds to a circularly polarized pulse.} \label{fig5}
\end{figure}

The HHG spectra for  monolayer and bilayer graphenes corresponding 
to different polarization of the driving laser pulse are shown in Fig.~\ref{fig5}.
The vector potential corresponding 
to the elliptically polarized pulse with an ellipticity $\epsilon$ is defined as
\begin{equation}
\textbf{A}(t) = \frac{A_0f(t)}{\sqrt{1+\epsilon^2}}\left[\cos(\omega t)\hat{\textbf{e}}_x + \epsilon \sin(\omega t)\hat{\textbf{e}}_y\right].
\end{equation}
Here, the same laser parameters are used as mentioned in the Section~\ref{section:2}. 
Both monolayer and bilayer graphenes show significant ellipticity-dependence in the harmonic yield.  A negligible harmonic yield is obtained for circularly polarized laser pulse. This indicates that using a single colour mid-infrared circular driver is not an appropriate 
method to generate circularly polarized harmonics from graphene. This has already been  experimentally demonstrated~\cite{taucer2017nonperturbative,yoshikawa2017high}. 
Recent theoretical studies revealed that efficient generation of circularly polarized harmonics is possible from graphene either by using a near-infrared circular laser pulse~\cite{chen2019circularly} or by using mid-infrared bi-circular counter-rotating laser pulses~\cite{jimenez2020light}. 
The harmonic spectrum corresponding to bilayer graphene shows the (6n$\pm$1) harmonic orders for circularly polarized laser, as expected from the symmetry considerations~\cite{gupta2003generation}. To have a better understanding about the variation of  the harmonics as a function of ellipticity of the driving laser, we show the integrated harmonic yield below.

\begin{figure}[h!]
	\includegraphics[width=\linewidth]{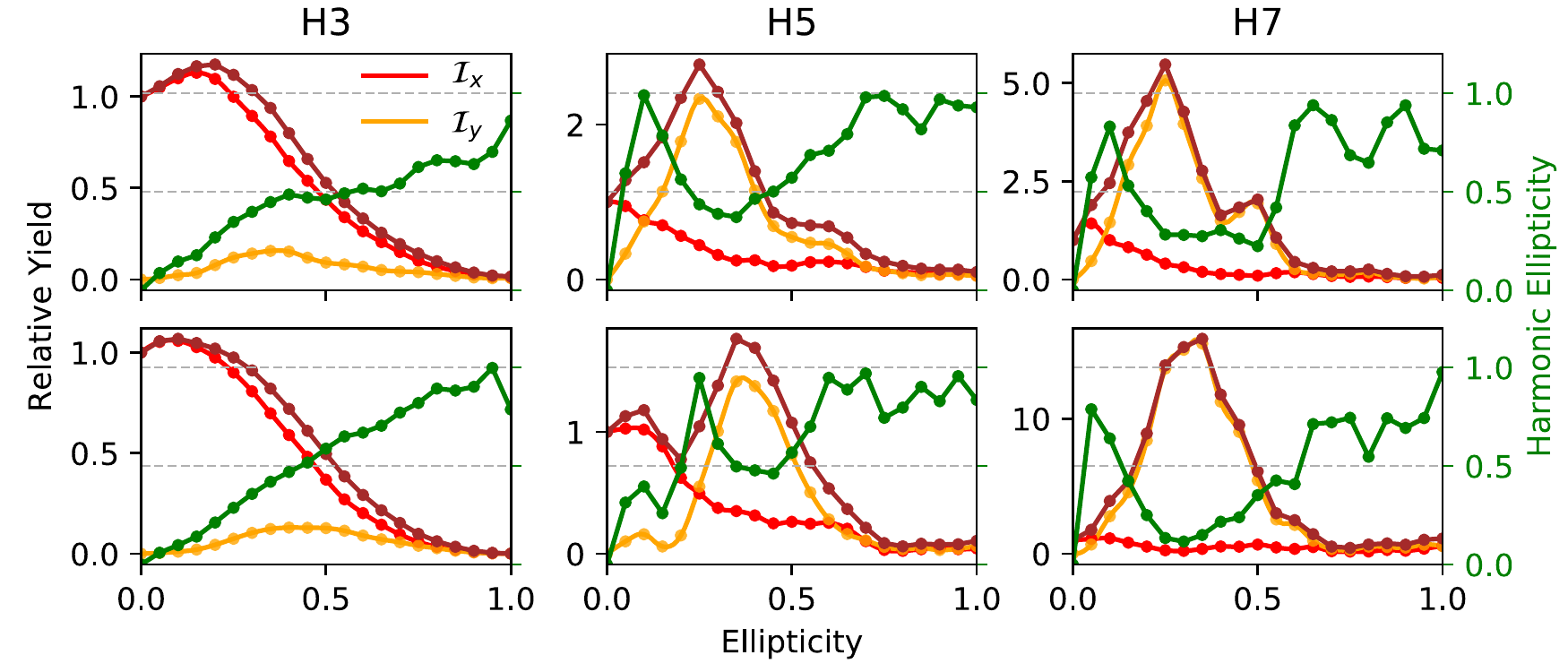}
	\caption{ Ellipticity dependence of the  integrated harmonic yield for 3$^{\textrm{rd}}$ (H3), 5$^{\textrm{th}}$ (H5), and 7$^{\textrm{th}}$ (H7) harmonics of the monolayer (top panel) and bilayer graphenes (bottom panel). 
	The integrated harmonic yield, and its $x$ and $y$-components are shown by brown, red and orange colour lines, respectively. The green line shows the ellipticity dependence on the averaged ellipticity of the emitted harmonics. 
	An elliptically polarized pulse with an intensity of 1$\times$10$^{11}$ W/cm$^2$ is used for HHG. } \label{fig4}
\end{figure}

The harmonic yield as a function of ellipticity for the monolayer (top panel) and bilayer graphenes (bottom panel) are presented in Fig.~\ref{fig4}.
The total harmonic yield is normalized with respect to the  harmonic yield for $\epsilon$ = 0. 
The ellipticity dependence of all the three harmonics agrees qualitatively well for monolayer and bilayer graphenes. The atomic-like ellipticity dependence of H3 can be attributed to its isotropic nature [see first column of Fig.~\ref{fig4}]. 
However, H5 and H7 show a characteristic ellipticity dependence. 
The harmonic yield has a maximum for a finite value of the ellipticity and is polarized along the normal direction of the major axis of the ellipse. 
This interesting feature was observed for the monolayer graphene experimentally 
and explained as a consequence of the semi-metallic nature of the monolayer graphene~\cite{yoshikawa2017high}. Since bilayer graphene is also semi-metallic,  it is also expected to exhibit similar ellipticity dependence, 
which we confirm here. 

The different qualitative behaviours of the ellipticity dependence of H3 compared to H5 and H7 
are also consistent with the findings that the interband and intraband mechanisms respond differently to the ellipticity of the driving laser~\cite{tancogne2017ellipticity} [see also insets of Figs.~\ref{fig2}(b) and 2(c)]. The characteristic ellipticity dependence of monolayer graphene was shown to be dominated by interband contributions in Ref.~\cite{liu2018driving}. The ellipticity of the maximum yield is different for bilayer graphene as a consequence of interlayer coupling.

The averaged ellipticity of the emitted harmonics as a function of the laser ellipticity shows interesting behaviour as  shown in Fig.~\ref{fig4} (see green colour).   
The averaged ellipticity of H3 of monolayer and bilayer graphene shows monotonically increasing behaviour. On the other hand,  
the behaviour is highly nonlinear for harmonics higher than H3. It is also interesting to note that harmonics with higher ellipticity can be obtained by a nearly linearly polarized pulse ($\epsilon <$0.3).

\section{Conclusions}

In summary,  HHG from monolayer and bilayer graphenes having AA and AB stacking is discussed. 
The harmonic spectra of the monolayer and bilayer graphenes are significantly different and exhibit 
characteristic features of having a vanishing band-gap. 
The HHG spectrum of the bilayer graphene shows signatures of the interlayer coupling, which affects 
high-order harmonics non-linearly and different harmonics are affected differently. 
The role of interlayer coupling was also found to be stacking dependent, resulting in the similar harmonic spectrum for monolayer and bilayer graphenes with AA stacking.
A strong interplay of the 
interband and intraband contributions to the total harmonic spectrum is noticed. 
Moreover,  characteristic polarization and ellipticity dependence are observed  in monolayer and bilayer graphenes.

\section{Acknowledgment}
G. D. acknowledges support from Science and Engineering Research Board (SERB) India 
(Project No. ECR/2017/001460).


\begin{thebibliography}{59}
\expandafter\ifx\csname natexlab\endcsname\relax\def\natexlab#1{#1}\fi
\expandafter\ifx\csname bibnamefont\endcsname\relax
  \def\bibnamefont#1{#1}\fi
\expandafter\ifx\csname bibfnamefont\endcsname\relax
  \def\bibfnamefont#1{#1}\fi
\expandafter\ifx\csname citenamefont\endcsname\relax
  \def\citenamefont#1{#1}\fi
\expandafter\ifx\csname url\endcsname\relax
  \def\url#1{\texttt{#1}}\fi
\expandafter\ifx\csname urlprefix\endcsname\relax\def\urlprefix{URL }\fi
\providecommand{\bibinfo}[2]{#2}
\providecommand{\eprint}[2][]{\url{#2}}

\bibitem[{\citenamefont{Kruchinin et~al.}(2018)\citenamefont{Kruchinin, Krausz,
  and Yakovlev}}]{kruchinin2018colloquium}
\bibinfo{author}{\bibfnamefont{S.~Y.} \bibnamefont{Kruchinin}},
  \bibinfo{author}{\bibfnamefont{F.}~\bibnamefont{Krausz}}, \bibnamefont{and}
  \bibinfo{author}{\bibfnamefont{V.~S.} \bibnamefont{Yakovlev}},
  \bibinfo{journal}{Reviews of Modern Physics} \textbf{\bibinfo{volume}{90}},
  \bibinfo{pages}{021002} (\bibinfo{year}{2018}).

\bibitem[{\citenamefont{Ghimire et~al.}(2011)\citenamefont{Ghimire, DiChiara,
  Sistrunk, Agostini, DiMauro, and Reis}}]{ghimire2011observation}
\bibinfo{author}{\bibfnamefont{S.}~\bibnamefont{Ghimire}},
  \bibinfo{author}{\bibfnamefont{A.~D.} \bibnamefont{DiChiara}},
  \bibinfo{author}{\bibfnamefont{E.}~\bibnamefont{Sistrunk}},
  \bibinfo{author}{\bibfnamefont{P.}~\bibnamefont{Agostini}},
  \bibinfo{author}{\bibfnamefont{L.~F.} \bibnamefont{DiMauro}},
  \bibnamefont{and} \bibinfo{author}{\bibfnamefont{D.~A.} \bibnamefont{Reis}},
  \bibinfo{journal}{Nature Physics} \textbf{\bibinfo{volume}{7}},
  \bibinfo{pages}{138} (\bibinfo{year}{2011}).

\bibitem[{\citenamefont{Ghimire and Reis}(2019)}]{ghimire2019}
\bibinfo{author}{\bibfnamefont{S.}~\bibnamefont{Ghimire}} \bibnamefont{and}
  \bibinfo{author}{\bibfnamefont{D.~A.} \bibnamefont{Reis}},
  \bibinfo{journal}{Nature Physics} \textbf{\bibinfo{volume}{15}},
  \bibinfo{pages}{10} (\bibinfo{year}{2019}).

\bibitem[{\citenamefont{Catoire et~al.}(2018)\citenamefont{Catoire, Bachau,
  Wang, Blaga, Agostini, and DiMauro}}]{catoire2018wannier}
\bibinfo{author}{\bibfnamefont{F.}~\bibnamefont{Catoire}},
  \bibinfo{author}{\bibfnamefont{H.}~\bibnamefont{Bachau}},
  \bibinfo{author}{\bibfnamefont{Z.}~\bibnamefont{Wang}},
  \bibinfo{author}{\bibfnamefont{C.}~\bibnamefont{Blaga}},
  \bibinfo{author}{\bibfnamefont{P.}~\bibnamefont{Agostini}}, \bibnamefont{and}
  \bibinfo{author}{\bibfnamefont{L.}~\bibnamefont{DiMauro}},
  \bibinfo{journal}{Physical review letters} \textbf{\bibinfo{volume}{121}},
  \bibinfo{pages}{143902} (\bibinfo{year}{2018}).

\bibitem[{\citenamefont{Vampa et~al.}(2015)\citenamefont{Vampa, Hammond,
  Thir{\'e}, Schmidt, L{\'e}gar{\'e}, McDonald, Brabec, Klug, and
  Corkum}}]{vampa2015all}
\bibinfo{author}{\bibfnamefont{G.}~\bibnamefont{Vampa}},
  \bibinfo{author}{\bibfnamefont{T.~J.} \bibnamefont{Hammond}},
  \bibinfo{author}{\bibfnamefont{N.}~\bibnamefont{Thir{\'e}}},
  \bibinfo{author}{\bibfnamefont{B.~E.} \bibnamefont{Schmidt}},
  \bibinfo{author}{\bibfnamefont{F.}~\bibnamefont{L{\'e}gar{\'e}}},
  \bibinfo{author}{\bibfnamefont{C.~R.} \bibnamefont{McDonald}},
  \bibinfo{author}{\bibfnamefont{T.}~\bibnamefont{Brabec}},
  \bibinfo{author}{\bibfnamefont{D.~D.} \bibnamefont{Klug}}, \bibnamefont{and}
  \bibinfo{author}{\bibfnamefont{P.~B.} \bibnamefont{Corkum}},
  \bibinfo{journal}{Physical Review Letters} \textbf{\bibinfo{volume}{115}},
  \bibinfo{pages}{193603} (\bibinfo{year}{2015}).

\bibitem[{\citenamefont{Lanin et~al.}(2017)\citenamefont{Lanin, Stepanov,
  Fedotov, and Zheltikov}}]{lanin2017mapping}
\bibinfo{author}{\bibfnamefont{A.~A.} \bibnamefont{Lanin}},
  \bibinfo{author}{\bibfnamefont{E.~A.} \bibnamefont{Stepanov}},
  \bibinfo{author}{\bibfnamefont{A.~B.} \bibnamefont{Fedotov}},
  \bibnamefont{and} \bibinfo{author}{\bibfnamefont{A.~M.}
  \bibnamefont{Zheltikov}}, \bibinfo{journal}{Optica}
  \textbf{\bibinfo{volume}{4}}, \bibinfo{pages}{516} (\bibinfo{year}{2017}).

\bibitem[{\citenamefont{Tancogne-Dejean
  et~al.}(2017{\natexlab{a}})\citenamefont{Tancogne-Dejean, M{\"u}cke,
  K{\"a}rtner, and Rubio}}]{tancogne2017impact}
\bibinfo{author}{\bibfnamefont{N.}~\bibnamefont{Tancogne-Dejean}},
  \bibinfo{author}{\bibfnamefont{O.~D.} \bibnamefont{M{\"u}cke}},
  \bibinfo{author}{\bibfnamefont{F.~X.} \bibnamefont{K{\"a}rtner}},
  \bibnamefont{and} \bibinfo{author}{\bibfnamefont{A.}~\bibnamefont{Rubio}},
  \bibinfo{journal}{Physical Review Letters} \textbf{\bibinfo{volume}{118}},
  \bibinfo{pages}{087403} (\bibinfo{year}{2017}{\natexlab{a}}).

\bibitem[{\citenamefont{Mrudul et~al.}(2020{\natexlab{a}})\citenamefont{Mrudul,
  Tancogne-Dejean, Rubio, and Dixit}}]{mrudul2020high}
\bibinfo{author}{\bibfnamefont{M.~S.} \bibnamefont{Mrudul}},
  \bibinfo{author}{\bibfnamefont{N.}~\bibnamefont{Tancogne-Dejean}},
  \bibinfo{author}{\bibfnamefont{A.}~\bibnamefont{Rubio}}, \bibnamefont{and}
  \bibinfo{author}{\bibfnamefont{G.}~\bibnamefont{Dixit}},
  \bibinfo{journal}{npj Computational Materials} \textbf{\bibinfo{volume}{6}},
  \bibinfo{pages}{1} (\bibinfo{year}{2020}{\natexlab{a}}).

\bibitem[{\citenamefont{Pattanayak et~al.}(2020)\citenamefont{Pattanayak,
  Mrudul, and Dixit}}]{pattanayak2020influence}
\bibinfo{author}{\bibfnamefont{A.}~\bibnamefont{Pattanayak}},
  \bibinfo{author}{\bibfnamefont{M.~S.} \bibnamefont{Mrudul}},
  \bibnamefont{and} \bibinfo{author}{\bibfnamefont{G.}~\bibnamefont{Dixit}},
  \bibinfo{journal}{Physical Review A} \textbf{\bibinfo{volume}{101}},
  \bibinfo{pages}{013404} (\bibinfo{year}{2020}).

\bibitem[{\citenamefont{Luu et~al.}(2015)\citenamefont{Luu, Garg, Kruchinin,
  Moulet, Hassan, and Goulielmakis}}]{luu2015extreme}
\bibinfo{author}{\bibfnamefont{T.~T.} \bibnamefont{Luu}},
  \bibinfo{author}{\bibfnamefont{M.}~\bibnamefont{Garg}},
  \bibinfo{author}{\bibfnamefont{S.~Y.} \bibnamefont{Kruchinin}},
  \bibinfo{author}{\bibfnamefont{A.}~\bibnamefont{Moulet}},
  \bibinfo{author}{\bibfnamefont{M.~T.} \bibnamefont{Hassan}},
  \bibnamefont{and}
  \bibinfo{author}{\bibfnamefont{E.}~\bibnamefont{Goulielmakis}},
  \bibinfo{journal}{Nature} \textbf{\bibinfo{volume}{521}},
  \bibinfo{pages}{498} (\bibinfo{year}{2015}).

\bibitem[{\citenamefont{Garg et~al.}(2016)\citenamefont{Garg, Zhan, Luu,
  Lakhotia, Klostermann, Guggenmos, and Goulielmakis}}]{garg2016multi}
\bibinfo{author}{\bibfnamefont{M.}~\bibnamefont{Garg}},
  \bibinfo{author}{\bibfnamefont{M.}~\bibnamefont{Zhan}},
  \bibinfo{author}{\bibfnamefont{T.~T.} \bibnamefont{Luu}},
  \bibinfo{author}{\bibfnamefont{H.}~\bibnamefont{Lakhotia}},
  \bibinfo{author}{\bibfnamefont{T.}~\bibnamefont{Klostermann}},
  \bibinfo{author}{\bibfnamefont{A.}~\bibnamefont{Guggenmos}},
  \bibnamefont{and}
  \bibinfo{author}{\bibfnamefont{E.}~\bibnamefont{Goulielmakis}},
  \bibinfo{journal}{Nature} \textbf{\bibinfo{volume}{538}},
  \bibinfo{pages}{359} (\bibinfo{year}{2016}).

\bibitem[{\citenamefont{Schubert et~al.}(2014)\citenamefont{Schubert,
  Hohenleutner, Langer, Urbanek, Lange, Huttner, Golde, Meier, Kira, Koch
  et~al.}}]{schubert2014sub}
\bibinfo{author}{\bibfnamefont{O.}~\bibnamefont{Schubert}},
  \bibinfo{author}{\bibfnamefont{M.}~\bibnamefont{Hohenleutner}},
  \bibinfo{author}{\bibfnamefont{F.}~\bibnamefont{Langer}},
  \bibinfo{author}{\bibfnamefont{B.}~\bibnamefont{Urbanek}},
  \bibinfo{author}{\bibfnamefont{C.}~\bibnamefont{Lange}},
  \bibinfo{author}{\bibfnamefont{U.}~\bibnamefont{Huttner}},
  \bibinfo{author}{\bibfnamefont{D.}~\bibnamefont{Golde}},
  \bibinfo{author}{\bibfnamefont{T.}~\bibnamefont{Meier}},
  \bibinfo{author}{\bibfnamefont{M.}~\bibnamefont{Kira}},
  \bibinfo{author}{\bibfnamefont{S.~W.} \bibnamefont{Koch}},
  \bibnamefont{et~al.}, \bibinfo{journal}{Nature Photonics}
  \textbf{\bibinfo{volume}{8}}, \bibinfo{pages}{119} (\bibinfo{year}{2014}).

\bibitem[{\citenamefont{McDonald et~al.}(2015)\citenamefont{McDonald, Vampa,
  Corkum, and Brabec}}]{mcdonald2015interband}
\bibinfo{author}{\bibfnamefont{C.~R.} \bibnamefont{McDonald}},
  \bibinfo{author}{\bibfnamefont{G.}~\bibnamefont{Vampa}},
  \bibinfo{author}{\bibfnamefont{P.~B.} \bibnamefont{Corkum}},
  \bibnamefont{and} \bibinfo{author}{\bibfnamefont{T.}~\bibnamefont{Brabec}},
  \bibinfo{journal}{Physical Review A} \textbf{\bibinfo{volume}{92}},
  \bibinfo{pages}{033845} (\bibinfo{year}{2015}).

\bibitem[{\citenamefont{Nourbakhsh et~al.}(2020)\citenamefont{Nourbakhsh,
  Tancogne-Dejean, Merdji, and Rubio}}]{nourbakhsh2020high}
\bibinfo{author}{\bibfnamefont{Z.}~\bibnamefont{Nourbakhsh}},
  \bibinfo{author}{\bibfnamefont{N.}~\bibnamefont{Tancogne-Dejean}},
  \bibinfo{author}{\bibfnamefont{H.}~\bibnamefont{Merdji}}, \bibnamefont{and}
  \bibinfo{author}{\bibfnamefont{A.}~\bibnamefont{Rubio}},
  \bibinfo{journal}{arXiv preprint arXiv:2010.08010}  (\bibinfo{year}{2020}).

\bibitem[{\citenamefont{Geim}(2009)}]{geim2009graphene}
\bibinfo{author}{\bibfnamefont{A.~K.} \bibnamefont{Geim}},
  \bibinfo{journal}{Science} \textbf{\bibinfo{volume}{324}},
  \bibinfo{pages}{1530} (\bibinfo{year}{2009}).

\bibitem[{\citenamefont{Novoselov et~al.}(2004)\citenamefont{Novoselov, Geim,
  Morozov, Jiang, Zhang, Dubonos, Grigorieva, and
  Firsov}}]{novoselov2004electric}
\bibinfo{author}{\bibfnamefont{K.~S.} \bibnamefont{Novoselov}},
  \bibinfo{author}{\bibfnamefont{A.~K.} \bibnamefont{Geim}},
  \bibinfo{author}{\bibfnamefont{S.~V.} \bibnamefont{Morozov}},
  \bibinfo{author}{\bibfnamefont{D.}~\bibnamefont{Jiang}},
  \bibinfo{author}{\bibfnamefont{Y.}~\bibnamefont{Zhang}},
  \bibinfo{author}{\bibfnamefont{S.~V.} \bibnamefont{Dubonos}},
  \bibinfo{author}{\bibfnamefont{I.~V.} \bibnamefont{Grigorieva}},
  \bibnamefont{and} \bibinfo{author}{\bibfnamefont{A.~A.}
  \bibnamefont{Firsov}}, \bibinfo{journal}{Science}
  \textbf{\bibinfo{volume}{306}}, \bibinfo{pages}{666} (\bibinfo{year}{2004}).

\bibitem[{\citenamefont{Avetissian and
  Mkrtchian}(2016)}]{avetissian2016coherent}
\bibinfo{author}{\bibfnamefont{H.~K.} \bibnamefont{Avetissian}}
  \bibnamefont{and} \bibinfo{author}{\bibfnamefont{G.~F.}
  \bibnamefont{Mkrtchian}}, \bibinfo{journal}{Physical Review B}
  \textbf{\bibinfo{volume}{94}}, \bibinfo{pages}{045419}
  (\bibinfo{year}{2016}).

\bibitem[{\citenamefont{Hendry et~al.}(2010)\citenamefont{Hendry, Hale, Moger,
  Savchenko, and Mikhailov}}]{hendry2010coherent}
\bibinfo{author}{\bibfnamefont{E.}~\bibnamefont{Hendry}},
  \bibinfo{author}{\bibfnamefont{P.~J.} \bibnamefont{Hale}},
  \bibinfo{author}{\bibfnamefont{J.}~\bibnamefont{Moger}},
  \bibinfo{author}{\bibfnamefont{A.~K.} \bibnamefont{Savchenko}},
  \bibnamefont{and} \bibinfo{author}{\bibfnamefont{S.~A.}
  \bibnamefont{Mikhailov}}, \bibinfo{journal}{Physical Review Letters}
  \textbf{\bibinfo{volume}{105}}, \bibinfo{pages}{097401}
  (\bibinfo{year}{2010}).

\bibitem[{\citenamefont{Kumar et~al.}(2013)\citenamefont{Kumar, Kumar,
  Gerstenkorn, Wang, Chiu, Smirl, and Zhao}}]{kumar2013third}
\bibinfo{author}{\bibfnamefont{N.}~\bibnamefont{Kumar}},
  \bibinfo{author}{\bibfnamefont{J.}~\bibnamefont{Kumar}},
  \bibinfo{author}{\bibfnamefont{C.}~\bibnamefont{Gerstenkorn}},
  \bibinfo{author}{\bibfnamefont{R.}~\bibnamefont{Wang}},
  \bibinfo{author}{\bibfnamefont{H.~Y.} \bibnamefont{Chiu}},
  \bibinfo{author}{\bibfnamefont{A.~L.} \bibnamefont{Smirl}}, \bibnamefont{and}
  \bibinfo{author}{\bibfnamefont{H.}~\bibnamefont{Zhao}},
  \bibinfo{journal}{Physical Review B} \textbf{\bibinfo{volume}{87}},
  \bibinfo{pages}{121406} (\bibinfo{year}{2013}).

\bibitem[{\citenamefont{Hafez et~al.}(2018)\citenamefont{Hafez, Kovalev,
  Deinert, Mics, Green, Awari, Chen, Germanskiy, Lehnert, Teichert
  et~al.}}]{hafez2018extremely}
\bibinfo{author}{\bibfnamefont{H.~A.} \bibnamefont{Hafez}},
  \bibinfo{author}{\bibfnamefont{S.}~\bibnamefont{Kovalev}},
  \bibinfo{author}{\bibfnamefont{J.-C.} \bibnamefont{Deinert}},
  \bibinfo{author}{\bibfnamefont{Z.}~\bibnamefont{Mics}},
  \bibinfo{author}{\bibfnamefont{B.}~\bibnamefont{Green}},
  \bibinfo{author}{\bibfnamefont{N.}~\bibnamefont{Awari}},
  \bibinfo{author}{\bibfnamefont{M.}~\bibnamefont{Chen}},
  \bibinfo{author}{\bibfnamefont{S.}~\bibnamefont{Germanskiy}},
  \bibinfo{author}{\bibfnamefont{U.}~\bibnamefont{Lehnert}},
  \bibinfo{author}{\bibfnamefont{J.}~\bibnamefont{Teichert}},
  \bibnamefont{et~al.}, \bibinfo{journal}{Nature}
  \textbf{\bibinfo{volume}{561}}, \bibinfo{pages}{507} (\bibinfo{year}{2018}).

\bibitem[{\citenamefont{Avetissian and Mkrtchian}(2018)}]{avetissian2018impact}
\bibinfo{author}{\bibfnamefont{H.}~\bibnamefont{Avetissian}} \bibnamefont{and}
  \bibinfo{author}{\bibfnamefont{G.}~\bibnamefont{Mkrtchian}},
  \bibinfo{journal}{Physical Review B} \textbf{\bibinfo{volume}{97}},
  \bibinfo{pages}{115454} (\bibinfo{year}{2018}).

\bibitem[{\citenamefont{Chizhova et~al.}(2017)\citenamefont{Chizhova, Libisch,
  and Burgd{\"o}rfer}}]{chizhova2017high}
\bibinfo{author}{\bibfnamefont{L.~A.} \bibnamefont{Chizhova}},
  \bibinfo{author}{\bibfnamefont{F.}~\bibnamefont{Libisch}}, \bibnamefont{and}
  \bibinfo{author}{\bibfnamefont{J.}~\bibnamefont{Burgd{\"o}rfer}},
  \bibinfo{journal}{Physical Review B} \textbf{\bibinfo{volume}{95}},
  \bibinfo{pages}{085436} (\bibinfo{year}{2017}).

\bibitem[{\citenamefont{Al-Naib et~al.}(2014)\citenamefont{Al-Naib, Sipe, and
  Dignam}}]{al2014high}
\bibinfo{author}{\bibfnamefont{I.}~\bibnamefont{Al-Naib}},
  \bibinfo{author}{\bibfnamefont{J.}~\bibnamefont{Sipe}}, \bibnamefont{and}
  \bibinfo{author}{\bibfnamefont{M.~M.} \bibnamefont{Dignam}},
  \bibinfo{journal}{Physical Review B} \textbf{\bibinfo{volume}{90}},
  \bibinfo{pages}{245423} (\bibinfo{year}{2014}).

\bibitem[{\citenamefont{Yoshikawa et~al.}(2017)\citenamefont{Yoshikawa, Tamaya,
  and Tanaka}}]{yoshikawa2017high}
\bibinfo{author}{\bibfnamefont{N.}~\bibnamefont{Yoshikawa}},
  \bibinfo{author}{\bibfnamefont{T.}~\bibnamefont{Tamaya}}, \bibnamefont{and}
  \bibinfo{author}{\bibfnamefont{K.}~\bibnamefont{Tanaka}},
  \bibinfo{journal}{Science} \textbf{\bibinfo{volume}{356}},
  \bibinfo{pages}{736} (\bibinfo{year}{2017}).

\bibitem[{\citenamefont{Zurr{\'o}n-Cifuentes
  et~al.}(2019)\citenamefont{Zurr{\'o}n-Cifuentes, Boyero-Garc{\'\i}a,
  Hern{\'a}ndez-Garc{\'\i}a, Pic{\'o}n, and Plaja}}]{zurron2019optical}
\bibinfo{author}{\bibfnamefont{{\'O}.}~\bibnamefont{Zurr{\'o}n-Cifuentes}},
  \bibinfo{author}{\bibfnamefont{R.}~\bibnamefont{Boyero-Garc{\'\i}a}},
  \bibinfo{author}{\bibfnamefont{C.}~\bibnamefont{Hern{\'a}ndez-Garc{\'\i}a}},
  \bibinfo{author}{\bibfnamefont{A.}~\bibnamefont{Pic{\'o}n}},
  \bibnamefont{and} \bibinfo{author}{\bibfnamefont{L.}~\bibnamefont{Plaja}},
  \bibinfo{journal}{Optics express} \textbf{\bibinfo{volume}{27}},
  \bibinfo{pages}{7776} (\bibinfo{year}{2019}).

\bibitem[{\citenamefont{Taucer et~al.}(2017)\citenamefont{Taucer, Hammond,
  Corkum, Vampa, Couture, Thir{\'e}, Schmidt, L{\'e}gar{\'e}, Selvi, Unsuree
  et~al.}}]{taucer2017nonperturbative}
\bibinfo{author}{\bibfnamefont{M.}~\bibnamefont{Taucer}},
  \bibinfo{author}{\bibfnamefont{T.}~\bibnamefont{Hammond}},
  \bibinfo{author}{\bibfnamefont{P.}~\bibnamefont{Corkum}},
  \bibinfo{author}{\bibfnamefont{G.}~\bibnamefont{Vampa}},
  \bibinfo{author}{\bibfnamefont{C.}~\bibnamefont{Couture}},
  \bibinfo{author}{\bibfnamefont{N.}~\bibnamefont{Thir{\'e}}},
  \bibinfo{author}{\bibfnamefont{B.}~\bibnamefont{Schmidt}},
  \bibinfo{author}{\bibfnamefont{F.}~\bibnamefont{L{\'e}gar{\'e}}},
  \bibinfo{author}{\bibfnamefont{H.}~\bibnamefont{Selvi}},
  \bibinfo{author}{\bibfnamefont{N.}~\bibnamefont{Unsuree}},
  \bibnamefont{et~al.}, \bibinfo{journal}{Physical Review B}
  \textbf{\bibinfo{volume}{96}}, \bibinfo{pages}{195420}
  (\bibinfo{year}{2017}).

\bibitem[{\citenamefont{Liu et~al.}(2018)\citenamefont{Liu, Zheng, Zeng, and
  Li}}]{liu2018driving}
\bibinfo{author}{\bibfnamefont{C.}~\bibnamefont{Liu}},
  \bibinfo{author}{\bibfnamefont{Y.}~\bibnamefont{Zheng}},
  \bibinfo{author}{\bibfnamefont{Z.}~\bibnamefont{Zeng}}, \bibnamefont{and}
  \bibinfo{author}{\bibfnamefont{R.}~\bibnamefont{Li}},
  \bibinfo{journal}{Physical Review A} \textbf{\bibinfo{volume}{97}},
  \bibinfo{pages}{063412} (\bibinfo{year}{2018}).

\bibitem[{\citenamefont{Chen and Qin}(2019)}]{chen2019circularly}
\bibinfo{author}{\bibfnamefont{Z.-Y.} \bibnamefont{Chen}} \bibnamefont{and}
  \bibinfo{author}{\bibfnamefont{R.}~\bibnamefont{Qin}},
  \bibinfo{journal}{Optics express} \textbf{\bibinfo{volume}{27}},
  \bibinfo{pages}{3761} (\bibinfo{year}{2019}).

\bibitem[{\citenamefont{Mikhailov}(2007)}]{mikhailov2007non}
\bibinfo{author}{\bibfnamefont{S.~A.} \bibnamefont{Mikhailov}},
  \bibinfo{journal}{EPL (Europhysics Letters)} \textbf{\bibinfo{volume}{79}},
  \bibinfo{pages}{27002} (\bibinfo{year}{2007}).

\bibitem[{\citenamefont{Gupta et~al.}(2003)\citenamefont{Gupta, Alon, and
  Moiseyev}}]{gupta2003generation}
\bibinfo{author}{\bibfnamefont{A.~K.} \bibnamefont{Gupta}},
  \bibinfo{author}{\bibfnamefont{O.~E.} \bibnamefont{Alon}}, \bibnamefont{and}
  \bibinfo{author}{\bibfnamefont{N.}~\bibnamefont{Moiseyev}},
  \bibinfo{journal}{Physical Review B} \textbf{\bibinfo{volume}{68}},
  \bibinfo{pages}{205101} (\bibinfo{year}{2003}).

\bibitem[{\citenamefont{Mrudul et~al.}(2020{\natexlab{b}})\citenamefont{Mrudul,
  Jim{\'e}nez-Gal{\'a}n, Ivanov, and Dixit}}]{jimenez2020light}
\bibinfo{author}{\bibfnamefont{M.~S.} \bibnamefont{Mrudul}},
  \bibinfo{author}{\bibfnamefont{{\'A}.}~\bibnamefont{Jim{\'e}nez-Gal{\'a}n}},
  \bibinfo{author}{\bibfnamefont{M.}~\bibnamefont{Ivanov}}, \bibnamefont{and}
  \bibinfo{author}{\bibfnamefont{G.}~\bibnamefont{Dixit}},
 \bibinfo{journal}{Optica} \textbf{\bibinfo{volume}{8}},
  \bibinfo{pages}{422--427} (\bibinfo{year}{2021}).

\bibitem[{\citenamefont{Sato et~al.}(2021)\citenamefont{Sato, Hirori, Sanari,
  Kanemitsu, and Rubio}}]{sato2021high}
\bibinfo{author}{\bibfnamefont{S.~A.} \bibnamefont{Sato}},
  \bibinfo{author}{\bibfnamefont{H.}~\bibnamefont{Hirori}},
  \bibinfo{author}{\bibfnamefont{Y.}~\bibnamefont{Sanari}},
  \bibinfo{author}{\bibfnamefont{Y.}~\bibnamefont{Kanemitsu}},
  \bibnamefont{and} \bibinfo{author}{\bibfnamefont{A.}~\bibnamefont{Rubio}},
  \bibinfo{journal}{Physical Review B} \textbf{\bibinfo{volume}{103}},
  \bibinfo{pages}{L041408} (\bibinfo{year}{2021}).

\bibitem[{\citenamefont{Zurr{\'o}n et~al.}(2018)\citenamefont{Zurr{\'o}n,
  Pic{\'o}n, and Plaja}}]{zurron2018theory}
\bibinfo{author}{\bibfnamefont{{\'O}.}~\bibnamefont{Zurr{\'o}n}},
  \bibinfo{author}{\bibfnamefont{A.}~\bibnamefont{Pic{\'o}n}},
  \bibnamefont{and} \bibinfo{author}{\bibfnamefont{L.}~\bibnamefont{Plaja}},
  \bibinfo{journal}{New Journal of Physics} \textbf{\bibinfo{volume}{20}},
  \bibinfo{pages}{053033} (\bibinfo{year}{2018}).

\bibitem[{\citenamefont{Vampa et~al.}(2014)\citenamefont{Vampa, McDonald,
  Orlando, Klug, Corkum, and Brabec}}]{vampa2014theoretical}
\bibinfo{author}{\bibfnamefont{G.}~\bibnamefont{Vampa}},
  \bibinfo{author}{\bibfnamefont{C.~R.} \bibnamefont{McDonald}},
  \bibinfo{author}{\bibfnamefont{G.}~\bibnamefont{Orlando}},
  \bibinfo{author}{\bibfnamefont{D.~D.} \bibnamefont{Klug}},
  \bibinfo{author}{\bibfnamefont{P.~B.} \bibnamefont{Corkum}},
  \bibnamefont{and} \bibinfo{author}{\bibfnamefont{T.}~\bibnamefont{Brabec}},
  \bibinfo{journal}{Physical Review Letters} \textbf{\bibinfo{volume}{113}},
  \bibinfo{pages}{073901} (\bibinfo{year}{2014}).

\bibitem[{\citenamefont{Kelardeh et~al.}(2015)\citenamefont{Kelardeh, Apalkov,
  and Stockman}}]{kelardeh2015graphene}
\bibinfo{author}{\bibfnamefont{H.~K.} \bibnamefont{Kelardeh}},
  \bibinfo{author}{\bibfnamefont{V.}~\bibnamefont{Apalkov}}, \bibnamefont{and}
  \bibinfo{author}{\bibfnamefont{M.~I.} \bibnamefont{Stockman}},
  \bibinfo{journal}{Physical Review B} \textbf{\bibinfo{volume}{91}},
  \bibinfo{pages}{045439} (\bibinfo{year}{2015}).

\bibitem[{\citenamefont{Tamaya et~al.}(2016)\citenamefont{Tamaya, Ishikawa,
  Ogawa, and Tanaka}}]{tamaya2016diabatic}
\bibinfo{author}{\bibfnamefont{T.}~\bibnamefont{Tamaya}},
  \bibinfo{author}{\bibfnamefont{A.}~\bibnamefont{Ishikawa}},
  \bibinfo{author}{\bibfnamefont{T.}~\bibnamefont{Ogawa}}, \bibnamefont{and}
  \bibinfo{author}{\bibfnamefont{K.}~\bibnamefont{Tanaka}},
  \bibinfo{journal}{Physical review letters} \textbf{\bibinfo{volume}{116}},
  \bibinfo{pages}{016601} (\bibinfo{year}{2016}).

\bibitem[{\citenamefont{Yan et~al.}(2012)\citenamefont{Yan, Li, Chandra,
  Tulevski, Wu, Freitag, Zhu, Avouris, and Xia}}]{yan2012tunable}
\bibinfo{author}{\bibfnamefont{H.}~\bibnamefont{Yan}},
  \bibinfo{author}{\bibfnamefont{X.}~\bibnamefont{Li}},
  \bibinfo{author}{\bibfnamefont{B.}~\bibnamefont{Chandra}},
  \bibinfo{author}{\bibfnamefont{G.}~\bibnamefont{Tulevski}},
  \bibinfo{author}{\bibfnamefont{Y.}~\bibnamefont{Wu}},
  \bibinfo{author}{\bibfnamefont{M.}~\bibnamefont{Freitag}},
  \bibinfo{author}{\bibfnamefont{W.}~\bibnamefont{Zhu}},
  \bibinfo{author}{\bibfnamefont{P.}~\bibnamefont{Avouris}}, \bibnamefont{and}
  \bibinfo{author}{\bibfnamefont{F.}~\bibnamefont{Xia}},
  \bibinfo{journal}{Nature Nanotechnology} \textbf{\bibinfo{volume}{7}},
  \bibinfo{pages}{330} (\bibinfo{year}{2012}).

\bibitem[{\citenamefont{Rozhkov et~al.}(2016)\citenamefont{Rozhkov, Sboychakov,
  Rakhmanov, and Nori}}]{rozhkov2016electronic}
\bibinfo{author}{\bibfnamefont{A.~V.} \bibnamefont{Rozhkov}},
  \bibinfo{author}{\bibfnamefont{A.~O.} \bibnamefont{Sboychakov}},
  \bibinfo{author}{\bibfnamefont{A.~L.} \bibnamefont{Rakhmanov}},
  \bibnamefont{and} \bibinfo{author}{\bibfnamefont{F.}~\bibnamefont{Nori}},
  \bibinfo{journal}{Physics Reports} \textbf{\bibinfo{volume}{648}},
  \bibinfo{pages}{1} (\bibinfo{year}{2016}).

\bibitem[{\citenamefont{McCann and Koshino}(2013)}]{mccann2013electronic}
\bibinfo{author}{\bibfnamefont{E.}~\bibnamefont{McCann}} \bibnamefont{and}
  \bibinfo{author}{\bibfnamefont{M.}~\bibnamefont{Koshino}},
  \bibinfo{journal}{Reports on Progress in Physics}
  \textbf{\bibinfo{volume}{76}}, \bibinfo{pages}{056503}
  (\bibinfo{year}{2013}).

\bibitem[{\citenamefont{Ulstrup et~al.}(2014)\citenamefont{Ulstrup, Johannsen,
  Cilento, Miwa, Crepaldi, Zacchigna, Cacho, Chapman, Springate, Mammadov
  et~al.}}]{ulstrup2014ultrafast}
\bibinfo{author}{\bibfnamefont{S.}~\bibnamefont{Ulstrup}},
  \bibinfo{author}{\bibfnamefont{J.~C.} \bibnamefont{Johannsen}},
  \bibinfo{author}{\bibfnamefont{F.}~\bibnamefont{Cilento}},
  \bibinfo{author}{\bibfnamefont{J.~A.} \bibnamefont{Miwa}},
  \bibinfo{author}{\bibfnamefont{A.}~\bibnamefont{Crepaldi}},
  \bibinfo{author}{\bibfnamefont{M.}~\bibnamefont{Zacchigna}},
  \bibinfo{author}{\bibfnamefont{C.}~\bibnamefont{Cacho}},
  \bibinfo{author}{\bibfnamefont{R.}~\bibnamefont{Chapman}},
  \bibinfo{author}{\bibfnamefont{E.}~\bibnamefont{Springate}},
  \bibinfo{author}{\bibfnamefont{S.}~\bibnamefont{Mammadov}},
  \bibnamefont{et~al.}, \bibinfo{journal}{Physical review letters}
  \textbf{\bibinfo{volume}{112}}, \bibinfo{pages}{257401}
  (\bibinfo{year}{2014}).

\bibitem[{\citenamefont{Kumar et~al.}(2020{\natexlab{a}})\citenamefont{Kumar,
  Herath, Apalkov, and Stockman}}]{kumar2020bilayer}
\bibinfo{author}{\bibfnamefont{P.}~\bibnamefont{Kumar}},
  \bibinfo{author}{\bibfnamefont{T.~M.} \bibnamefont{Herath}},
  \bibinfo{author}{\bibfnamefont{V.}~\bibnamefont{Apalkov}}, \bibnamefont{and}
  \bibinfo{author}{\bibfnamefont{M.~I.} \bibnamefont{Stockman}},
  \bibinfo{journal}{arXiv preprint arXiv:2004.09732}
  (\bibinfo{year}{2020}{\natexlab{a}}).

\bibitem[{\citenamefont{Hipolito et~al.}(2018)\citenamefont{Hipolito,
  Taghizadeh, and Pedersen}}]{hipolito2018nonlinear}
\bibinfo{author}{\bibfnamefont{F.}~\bibnamefont{Hipolito}},
  \bibinfo{author}{\bibfnamefont{A.}~\bibnamefont{Taghizadeh}},
  \bibnamefont{and} \bibinfo{author}{\bibfnamefont{T.~G.}
  \bibnamefont{Pedersen}}, \bibinfo{journal}{Physical Review B}
  \textbf{\bibinfo{volume}{98}}, \bibinfo{pages}{205420}
  (\bibinfo{year}{2018}).

\bibitem[{\citenamefont{Kumar et~al.}(2020{\natexlab{b}})\citenamefont{Kumar,
  Herath, Apalkov, and Stockman}}]{kumar2020ultrafast}
\bibinfo{author}{\bibfnamefont{P.}~\bibnamefont{Kumar}},
  \bibinfo{author}{\bibfnamefont{T.~M.} \bibnamefont{Herath}},
  \bibinfo{author}{\bibfnamefont{V.}~\bibnamefont{Apalkov}}, \bibnamefont{and}
  \bibinfo{author}{\bibfnamefont{M.~I.} \bibnamefont{Stockman}},
  \bibinfo{journal}{arXiv preprint arXiv:2007.13480}
  (\bibinfo{year}{2020}{\natexlab{b}}).

\bibitem[{\citenamefont{Avetissian et~al.}(2013)\citenamefont{Avetissian,
  Mkrtchian, Batrakov, Maksimenko, and Hoffmann}}]{avetissian2013multiphoton}
\bibinfo{author}{\bibfnamefont{H.}~\bibnamefont{Avetissian}},
  \bibinfo{author}{\bibfnamefont{G.}~\bibnamefont{Mkrtchian}},
  \bibinfo{author}{\bibfnamefont{K.}~\bibnamefont{Batrakov}},
  \bibinfo{author}{\bibfnamefont{S.}~\bibnamefont{Maksimenko}},
  \bibnamefont{and} \bibinfo{author}{\bibfnamefont{A.}~\bibnamefont{Hoffmann}},
  \bibinfo{journal}{Physical Review B} \textbf{\bibinfo{volume}{88}},
  \bibinfo{pages}{165411} (\bibinfo{year}{2013}).

\bibitem[{\citenamefont{Ikeda}(2020)}]{PhysRevResearch.2.032015}
\bibinfo{author}{\bibfnamefont{T.~N.} \bibnamefont{Ikeda}},
  \bibinfo{journal}{Physical Review Research} \textbf{\bibinfo{volume}{2}},
  \bibinfo{pages}{032015} (\bibinfo{year}{2020}),
  \urlprefix\url{https://link.aps.org/doi/10.1103/PhysRevResearch.2.032015}.

\bibitem[{\citenamefont{Reich et~al.}(2002)\citenamefont{Reich, Maultzsch,
  Thomsen, and Ordejon}}]{reich2002tight}
\bibinfo{author}{\bibfnamefont{S.}~\bibnamefont{Reich}},
  \bibinfo{author}{\bibfnamefont{J.}~\bibnamefont{Maultzsch}},
  \bibinfo{author}{\bibfnamefont{C.}~\bibnamefont{Thomsen}}, \bibnamefont{and}
  \bibinfo{author}{\bibfnamefont{P.}~\bibnamefont{Ordejon}},
  \bibinfo{journal}{Physical Review B} \textbf{\bibinfo{volume}{66}},
  \bibinfo{pages}{035412} (\bibinfo{year}{2002}).

\bibitem[{\citenamefont{Trambly~de Laissardi{\`e}re
  et~al.}(2010)\citenamefont{Trambly~de Laissardi{\`e}re, Mayou, and
  Magaud}}]{trambly2010localization}
\bibinfo{author}{\bibfnamefont{G.}~\bibnamefont{Trambly~de Laissardi{\`e}re}},
  \bibinfo{author}{\bibfnamefont{D.}~\bibnamefont{Mayou}}, \bibnamefont{and}
  \bibinfo{author}{\bibfnamefont{L.}~\bibnamefont{Magaud}},
  \bibinfo{journal}{Nano letters} \textbf{\bibinfo{volume}{10}},
  \bibinfo{pages}{804} (\bibinfo{year}{2010}).

\bibitem[{\citenamefont{Moon and Koshino}(2012)}]{moon2012energy}
\bibinfo{author}{\bibfnamefont{P.}~\bibnamefont{Moon}} \bibnamefont{and}
  \bibinfo{author}{\bibfnamefont{M.}~\bibnamefont{Koshino}},
  \bibinfo{journal}{Physical Review B} \textbf{\bibinfo{volume}{85}},
  \bibinfo{pages}{195458} (\bibinfo{year}{2012}).

\bibitem[{\citenamefont{Houston}(1940)}]{houston1940acceleration}
\bibinfo{author}{\bibfnamefont{W.}~\bibnamefont{Houston}},
  \bibinfo{journal}{Physical Review} \textbf{\bibinfo{volume}{57}},
  \bibinfo{pages}{184} (\bibinfo{year}{1940}).

\bibitem[{\citenamefont{Krieger and Iafrate}(1986)}]{krieger1986time}
\bibinfo{author}{\bibfnamefont{J.}~\bibnamefont{Krieger}} \bibnamefont{and}
  \bibinfo{author}{\bibfnamefont{G.}~\bibnamefont{Iafrate}},
  \bibinfo{journal}{Physical Review B} \textbf{\bibinfo{volume}{33}},
  \bibinfo{pages}{5494} (\bibinfo{year}{1986}).

\bibitem[{\citenamefont{Floss et~al.}(2018)\citenamefont{Floss, Lemell,
  Wachter, Smejkal, Sato, Tong, Yabana, and Burgd{\"o}rfer}}]{floss2018ab}
\bibinfo{author}{\bibfnamefont{I.}~\bibnamefont{Floss}},
  \bibinfo{author}{\bibfnamefont{C.}~\bibnamefont{Lemell}},
  \bibinfo{author}{\bibfnamefont{G.}~\bibnamefont{Wachter}},
  \bibinfo{author}{\bibfnamefont{V.}~\bibnamefont{Smejkal}},
  \bibinfo{author}{\bibfnamefont{S.~A.} \bibnamefont{Sato}},
  \bibinfo{author}{\bibfnamefont{X.-M.} \bibnamefont{Tong}},
  \bibinfo{author}{\bibfnamefont{K.}~\bibnamefont{Yabana}}, \bibnamefont{and}
  \bibinfo{author}{\bibfnamefont{J.}~\bibnamefont{Burgd{\"o}rfer}},
  \bibinfo{journal}{Physical Review A} \textbf{\bibinfo{volume}{97}},
  \bibinfo{pages}{011401} (\bibinfo{year}{2018}).

\bibitem[{\citenamefont{Wu et~al.}(2015)\citenamefont{Wu, Ghimire, Reis,
  Schafer, and Gaarde}}]{wu2015high}
\bibinfo{author}{\bibfnamefont{M.}~\bibnamefont{Wu}},
  \bibinfo{author}{\bibfnamefont{S.}~\bibnamefont{Ghimire}},
  \bibinfo{author}{\bibfnamefont{D.~A.} \bibnamefont{Reis}},
  \bibinfo{author}{\bibfnamefont{K.~J.} \bibnamefont{Schafer}},
  \bibnamefont{and} \bibinfo{author}{\bibfnamefont{M.~B.}
  \bibnamefont{Gaarde}}, \bibinfo{journal}{Physical Review A}
  \textbf{\bibinfo{volume}{91}}, \bibinfo{pages}{043839}
  (\bibinfo{year}{2015}).

\bibitem[{\citenamefont{Stroucken et~al.}(2011)\citenamefont{Stroucken,
  Gr{\"o}nqvist, and Koch}}]{stroucken2011optical}
\bibinfo{author}{\bibfnamefont{T.}~\bibnamefont{Stroucken}},
  \bibinfo{author}{\bibfnamefont{J.}~\bibnamefont{Gr{\"o}nqvist}},
  \bibnamefont{and} \bibinfo{author}{\bibfnamefont{S.}~\bibnamefont{Koch}},
  \bibinfo{journal}{Physical Review B} \textbf{\bibinfo{volume}{84}},
  \bibinfo{pages}{205445} (\bibinfo{year}{2011}).

\bibitem[{\citenamefont{Charlier et~al.}(1991)\citenamefont{Charlier, Gonze,
  and Michenaud}}]{charlier1991first}
\bibinfo{author}{\bibfnamefont{J.-C.} \bibnamefont{Charlier}},
  \bibinfo{author}{\bibfnamefont{X.}~\bibnamefont{Gonze}}, \bibnamefont{and}
  \bibinfo{author}{\bibfnamefont{J.-P.} \bibnamefont{Michenaud}},
  \bibinfo{journal}{Physical Review B} \textbf{\bibinfo{volume}{43}},
  \bibinfo{pages}{4579} (\bibinfo{year}{1991}).

\bibitem[{\citenamefont{Min et~al.}(2007)\citenamefont{Min, Sahu, Banerjee, and
  MacDonald}}]{min2007ab}
\bibinfo{author}{\bibfnamefont{H.}~\bibnamefont{Min}},
  \bibinfo{author}{\bibfnamefont{B.}~\bibnamefont{Sahu}},
  \bibinfo{author}{\bibfnamefont{S.~K.} \bibnamefont{Banerjee}},
  \bibnamefont{and}
  \bibinfo{author}{\bibfnamefont{A.}~\bibnamefont{MacDonald}},
  \bibinfo{journal}{Physical Review B} \textbf{\bibinfo{volume}{75}},
  \bibinfo{pages}{155115} (\bibinfo{year}{2007}).

\bibitem[{\citenamefont{You et~al.}(2017)\citenamefont{You, Reis, and
  Ghimire}}]{you2017anisotropic}
\bibinfo{author}{\bibfnamefont{Y.~S.} \bibnamefont{You}},
  \bibinfo{author}{\bibfnamefont{D.~A.} \bibnamefont{Reis}}, \bibnamefont{and}
  \bibinfo{author}{\bibfnamefont{S.}~\bibnamefont{Ghimire}},
  \bibinfo{journal}{Nature physics} \textbf{\bibinfo{volume}{13}},
  \bibinfo{pages}{345} (\bibinfo{year}{2017}).

\bibitem[{\citenamefont{Mrudul et~al.}(2019)\citenamefont{Mrudul, Pattanayak,
  Ivanov, and Dixit}}]{pattanayak2019direct}
\bibinfo{author}{\bibfnamefont{M.~S.} \bibnamefont{Mrudul}},
  \bibinfo{author}{\bibfnamefont{A.}~\bibnamefont{Pattanayak}},
  \bibinfo{author}{\bibfnamefont{M.}~\bibnamefont{Ivanov}}, \bibnamefont{and}
  \bibinfo{author}{\bibfnamefont{G.}~\bibnamefont{Dixit}},
  \bibinfo{journal}{Physical Review A} \textbf{\bibinfo{volume}{100}},
  \bibinfo{pages}{043420} (\bibinfo{year}{2019}).

\bibitem[{\citenamefont{Parks et~al.}(2020)\citenamefont{Parks, Ernotte,
  Thorpe, McDonald, Corkum, Taucer, and Brabec}}]{parks2020wannier}
\bibinfo{author}{\bibfnamefont{A.~M.} \bibnamefont{Parks}},
  \bibinfo{author}{\bibfnamefont{G.}~\bibnamefont{Ernotte}},
  \bibinfo{author}{\bibfnamefont{A.}~\bibnamefont{Thorpe}},
  \bibinfo{author}{\bibfnamefont{C.~R.} \bibnamefont{McDonald}},
  \bibinfo{author}{\bibfnamefont{P.~B.} \bibnamefont{Corkum}},
  \bibinfo{author}{\bibfnamefont{M.}~\bibnamefont{Taucer}}, \bibnamefont{and}
  \bibinfo{author}{\bibfnamefont{T.}~\bibnamefont{Brabec}},
  \bibinfo{journal}{arXiv preprint arXiv:2006.09651}  (\bibinfo{year}{2020}).

\bibitem[{\citenamefont{Tancogne-Dejean
  et~al.}(2017{\natexlab{b}})\citenamefont{Tancogne-Dejean, M{\"u}cke,
  K{\"a}rtner, and Rubio}}]{tancogne2017ellipticity}
\bibinfo{author}{\bibfnamefont{N.}~\bibnamefont{Tancogne-Dejean}},
  \bibinfo{author}{\bibfnamefont{O.~D.} \bibnamefont{M{\"u}cke}},
  \bibinfo{author}{\bibfnamefont{F.~X.} \bibnamefont{K{\"a}rtner}},
  \bibnamefont{and} \bibinfo{author}{\bibfnamefont{A.}~\bibnamefont{Rubio}},
  \bibinfo{journal}{Nature Communications} \textbf{\bibinfo{volume}{8}},
  \bibinfo{pages}{745} (\bibinfo{year}{2017}{\natexlab{b}}).

\end{thebibliography}

\end{document}